\documentclass[12pt]{article}

\usepackage{amstext}
\usepackage{latexsym}
\usepackage{amssymb}
\usepackage{amsmath}

\def\R{\mathbb R}

\begin{document}

\newtheorem{theorem}{Theorem}[section]
\renewcommand{\thetheorem}{\arabic{section}.\arabic{theorem}}
\newtheorem{definition}[theorem]{Definition}
\newtheorem{deflem}[theorem]{Definition and Lemma}
\newtheorem{lemma}[theorem]{Lemma}
\newtheorem{example}[theorem]{Example}

\newtheorem{remark}[theorem]{Remark}
\newtheorem{remarks}[theorem]{Remarks}
\newtheorem{cor}[theorem]{Corollary}
\newtheorem{pro}[theorem]{Proposition}
\newtheorem{proposition}[theorem]{Proposition}
\newtheorem{assumption}[theorem]{Assumption}

\renewcommand{\theequation}{\thesection.\arabic{equation}}

\title{Multipole Radiation in a Collisionless Gas Coupled \\ 
to Electromagnetism or Scalar Gravitation}
\author{{\sc S.~Bauer\footnote{Supported in parts by
        DFG priority research program SPP 1095}$\,\,^{1}$,
        M.~Kunze$^{1}$, G.~Rein$^{2}$\,\, \& A.~D.~Rendall$^{3}$} \\[2ex]
        $^{1}$ Universit\"at Duisburg-Essen, Fachbereich Mathematik, \\
        D\,-\,45117 Essen, Germany \\[1ex]
        $^{2}$ Universit\"at Bayreuth, Fakult\"at f\"ur Mathematik und Physik, \\
        D-95440 Bayreuth, Germany \\[1ex]
        $^{3}$ Max-Planck-Institut f\"ur Gravitationsphysik, Am M\"uhlenberg 1, \\
        D\,-\,14476 Golm, Germany \\[2ex]}
\date{}
\maketitle

\begin{abstract}\noindent 
We consider the relativistic Vlasov-Maxwell and Vlasov-Nordstr\"om systems 
which describe large particle ensembles interacting by either
electromagnetic fields or a relativistic scalar gravity model. 
For both systems we derive a radiation formula analogous to the
Einstein quadrupole formula in general relativity.
\end{abstract}


\setcounter{equation}{0}

\section{Introduction and Main Results}

This paper is an investigation of the mathematical properties of certain
models for the interaction of matter, described by a kinetic equation,
with radiation, described by hyperbolic equations. The first model,
the relativistic Vlasov-Maxwell system, plays an important role in plasma
physics. The motivation for studying the second model, the 
Vlasov-Nordstr\"om system, comes from the theory of gravitation.
On a mathematical level the Vlasov-Maxwell system can also give insights
into gravity.

The most precise existing theory of gravitation, general relativity,
predicts that certain astrophysical systems, such as colliding
black holes or neutron stars, will give rise to gravitational radiation.
There is a major international effort under way to detect these
gravitational waves \cite{Gravwaves}. In order to relate the general theory
to predictions of what the detectors will see it is necessary to use
approximation methods - the exact theory is too complicated. The
mathematical status of these approximations remains unclear although
partial results exist. This paper is intended as a contribution to
understanding the mathematical structures involved.

Since the solutions of the equations of general relativity are so
difficult to analyze rigorously it is useful to start with model
problems. One possibility is the scalar theory of gravitation
considered here, the Vlasov-Nordstr\"om theory \cite{calo1}.
It has already been used as a model problem for numerical relativity
in \cite{shapi}.

Among the approximation methods used to study gravitational radiation
those which are most accessible mathematically are the
post-Newtonian approximations. Some information on these has been
obtained in \cite{ADR1} and \cite{ADR2}. Results which are analogous to these
but go much further have been obtained for the Vlasov-Maxwell
and Vlasov-Nordstr\"om systems in \cite{bauku} and \cite{bauer} respectively.
None of these results include radiation explicitly. Here we take a first
step in doing so. On the other hand, for the case of finite particle
systems interacting with their self-induced fields there are several
rigorous results concerning radiation; see \cite{spohn} 
for an up-to-date review.

Our main results (Theorem~\ref{main2} and Theorem~\ref{main-gen} below) 
are relations between the motion of matter and the radiation flux at infinity
for the Vlasov-Maxwell and Vlasov-Nordstr\"om systems respectively. They
are analogues of the Einstein quadrupole formula \cite[(4.5.13)]{straum} which
is a basic tool in computing the flux of gravitational waves from
a given source. In the case of the Einstein and Maxwell
equations a spherically symmetric system does not radiate. For
the Vlasov-Nordstr\"om system a spherical system can radiate
and the specialization of the general formula to that case is computed.
In \cite{shapi} a difference between the spherically symmetric and the
general case was claimed but we have not succeeded in connecting
this to our results. The main theorems are obtained under plausible
assumptions on the behavior of global solutions of the relevant
system (Assumption~\ref{ex-assum-VM} and Assumption~\ref{ex-assum} below).
The former can be proved to hold in the case of small data.

For the systems we are going to consider the (scalar) energy density $e$
and the (vector) momentum density ${\cal P}$ are related by the conservation law
\[ \partial_t e+\nabla\cdot {\cal P}=0. \]
Defining the local energy in the ball of radius $r>0$ as
\[ {\cal E}_r(t)=\int_{|x|\le r} e(t, x)\,dx, \]
this conservation law and the divergence theorem imply that
\begin{equation}\label{ddtE-form}
   \frac{d}{dt}\,{\cal E}_r(t)=\int_{|x|\le r}\partial_t e(t, x)\,dx
   =-\int_{|x|\le r} \nabla\cdot {\cal P}(t, x)\,dx
   =-\int_{|x|=r} \bar{x}\cdot {\cal P}(t, x)\,d\sigma(x),
\end{equation}
where $\bar{x}=\frac{x}{|x|}$ denotes the outer unit normal.
More specifically, for the relativistic Vlasov-Maxwell system with two particle species,
\begin{eqnarray}
  e_{{\rm RVM}}(t, x) & = & c^2\int\sqrt{1+c^{-2}p^2}\,(f^+ + f^-)(t, x, p)\,dp
  +\frac{1}{8\pi}\big(|E(t, x)|^2+|B(t,x)|^2\big), \label{ervm-def} \\
  {\cal P}_{{\rm RVM}}(t, x) & = & c^2\int p (f^++f^-)(t, x, p)\,dp
  +\frac{c}{4\pi}\,E(t, x)\times B(t, x), \label{Prvm-def}
\end{eqnarray}
whereas for the Vlasov-Nordstr\"om system,
\begin{eqnarray}
  e_{{\rm VN}}(t, x) & = & c^2\int\sqrt{1+c^{-2}p^2}\,f(t, x, p)\,dp
  +\frac{c^2}{8\pi}\big((\partial_t\phi(t, x))^2+c^2|\nabla\phi(t, x)|^2\big), \label{evn-def}
  \\ {\cal P}_{{\rm VN}}(t, x) & = & c^2\int p f(t, x, p)\,dp
  -\frac{c^4}{4\pi}\,\partial_t\phi(t, x)\,\nabla\phi(t, x) . \nonumber
\end{eqnarray}
Our assumptions on the support
of the distribution function
will be such that the contributions of $\int p (f^+ + f^-)\,dp$
to ${\cal P}_{{\rm RVM}}$ and $\int p f\,dp$ to ${\cal P}_{{\rm VN}}$
vanish for $|x|=r$ large. Hence we arrive at
\[ \frac{d}{dt}\,{\cal E}^{{\rm RVM}}_r(t)=\frac{c}{4\pi}\int_{|x|=r}
   \bar{x}\cdot (B\times E)(t, x)\,d\sigma(x) \]
for the relativistic Vlasov-Maxwell system, and
\[ \frac{d}{dt}\,{\cal E}^{{\rm VN}}_r(t)=\frac{c^4}{4\pi}\int_{|x|=r}
   \bar{x}\cdot (\partial_t\phi\nabla\phi)(t, x)\,d\sigma(x) \]
for the Vlasov-Nordstr\"om system.

The main results of this paper are concerned with the expansion of these energy fluxes
for $r, c\to\infty$ and $|t-c^{-1}r|\le {\rm const}$.
Under suitable assumptions we will prove that, to leading order,
\[ \frac{d}{dt}\,{\cal E}^{{\rm RVM}}_r(t)
   \sim -\frac{2}{3c^3}\,|\partial_t^2 {\cal D}(u)|^2, \]
where $u=t-c^{-1}r$ denotes the retarded time and ${\cal D}(u)=\int x\,\rho_0(u, x)\,dx$
is the dipole moment associated to the Newtonian limit of the relativistic Vlasov-Maxwell system.
Similarly,
\[ \frac{d}{dt}\,{\cal E}^{{\rm VN}}_r(t)\sim -\frac{1}{4\pi c^5}\int_{|\omega|=1}
   \big(\partial_t {\cal R}(\omega, u)\big)^2\,d\sigma(\omega), \]
with a more complicated radiation term ${\cal R}$ associated to the Newtonian limit
of the Vlasov-Nordstr\"om system. In the spherically symmetric case,
$\partial_t {\cal R}(\omega, u)$ is found to be proportional to $\partial_t {\cal E}_{{\rm kin}}(u)$,
the change of kinetic energy of the Newtonian system. The exact statements are contained
in Theorems~\ref{main2} and \ref{main-gen} below.

\subsection{Dipole Radiation in the Relativistic Vlasov-Maxwell System}

The relativistic Vlasov-Maxwell system describes a large ensemble of particles
which move at possibly relativistic speeds and interact only by the electromagnetic fields
which the ensemble creates collectively.
Collisions among the particles are assumed to be sufficiently rare to be
neglected \cite{glas}. In order to see effects due to radiation damping
it is necessary that there are at least two species of particles with different
charge-to-mass ratios. For the sake of simplicity we assume that there
are exactly two species with their masses normalized to unity and their charges
normalized to plus and minus unity, respectively. The density of the positively
and negatively charged particles in phase space is given by
the non-negative distribution functions
$f^\pm=f^\pm(t, x, p)$, depending on time $t\in\R$, position $x\in\R^3$,
and momentum $p\in\R^3$. Their dynamics is governed by the relativistic Vlasov-Maxwell system
\begin{equation}\label{RVMc}
   \left.\begin{array}{c}
   \partial_t f^\pm+\hat{p}\cdot\nabla_x f^\pm
   \pm(E+c^{-1}\hat{p}\times B)\cdot\nabla_p f^\pm=0, \\[1ex]
   c\,\nabla\times E=-\partial_t B,\quad c\,\nabla\times B=\partial_t E+4\pi j, \\[1ex]
   \nabla\cdot E=4\pi\rho,\quad\nabla\cdot B=0, \\[1ex]
   \displaystyle\rho=\int (f^+-f^-)\,dp,\quad j=\int\hat{p}\,(f^+-f^-)\,dp,
   \end{array}\quad\right\}\tag{RVMc}
\end{equation}
where
\begin{equation}\label{hatp-def}
   \hat{p}=\gamma p,\quad\gamma=(1+c^{-2}p^2)^{-1/2},\quad p^2=|p|^2,
   \quad\mbox{and}\quad\int=\int_{\R^3}.
\end{equation}
The electric field $E=E(t, x)\in\R^3$ and the magnetic field $B=B(t, x)\in\R^3$
satisfy the wave equations
\begin{equation}\label{EB-wave}
   (-\partial_t^2+c^2\Delta)E=4\pi (c^2\nabla\rho+\partial_t j)
   \quad\mbox{and}\quad (-\partial_t^2+c^2\Delta)B=-4\pi c\,\nabla\times j.
\end{equation}
In order to determine the radiation of the system at infinity, we have to consider solutions
that are isolated from incoming radiation. For the wave equations in (\ref{EB-wave}),
this means that we need to restrict ourselves to the retarded part of the solutions.
Accordingly, (\ref{RVMc}) is replaced by
\begin{equation}\label{retRVMc}
   \left.\begin{array}{c}
   \partial_t f^\pm+\hat{p}\cdot\nabla_x f^\pm
   \pm(E+c^{-1}\hat{p}\times B)\cdot\nabla_p f^\pm=0, \\[1ex]
   \displaystyle E(t, x)=-\int
   (\nabla\rho+c^{-2}\partial_t j)(t-c^{-1}|y-x|, y)\,\frac{dy}{|y-x|}, \\[2.5ex]
   \displaystyle B(t, x)=c^{-1}\int
   \nabla\times j(t-c^{-1}|y-x|, y)\,\frac{dy}{|y-x|}, \\[2.5ex]
   \displaystyle\rho=\int (f^+-f^-)\,dp,\quad j=\int\hat{p}\,(f^+-f^-)\,dp,
   \end{array}\quad\right\}\tag{retRVMc}
\end{equation}
which we call the retarded relativistic Vlasov-Maxwell system.
We prescribe initial data
\begin{equation}\label{indat}
   f^\pm(0, x, p)=f^{\pm,\circ}(x, p),\quad x, p\in\R^3,
\end{equation}
for the densities at $t=0$; these data do not depend on $c$.
However, the corresponding solution $(f^+, f^-, E, B)$ does depend on $c$,
but we do not make explicit this dependence through our notation.
We refer to Remark~\ref{remVM}(c) below for the case of initial data varying with $c$.
Our standing assumption is that the initial data are non-negative,
smooth, and compactly supported,
\begin{equation}\label{daten1}
   f^{\pm,\circ}\in C^\infty_0 (\R^3 \times \R^3),\quad f^{\pm,\circ}\geq 0,
\end{equation}
and we fix positive constants $R_0, P_0, S_0$ such that
\begin{equation}\label{daten2}
   f^{\pm,\circ} (x,p) = 0\quad\mbox{for}\quad |x|\ge R_0\quad\mbox{or}\quad |p|\ge P_0,
   \quad\mbox{and}\quad {\|f^{\pm,\circ}\|}_{W^{3,\infty}}\le S_0.
\end{equation}
Every solution of (\ref{retRVMc}) satisfies the identity
\begin{equation}\label{fpmdarst}
   f^\pm(t, x, p) = f^{\pm, \circ} (X^\pm(0,t,x,p), P^\pm(0,t,x,p)), 
\end{equation}
where $s\mapsto (X^\pm(s, t, x, p), P^\pm(s, t, x, p))$ solves the characteristic system
\begin{equation}\label{olan}
   \dot x = \hat p,\quad\dot p = \pm(E+c^{-1}\hat{p}\times B),
\end{equation}
with data $X^\pm(t, t, x, p)=x$ and $P^\pm(t, t, x, p)=p$. Hence $0 \leq f^\pm(t,x,p)
\leq {\|f^{\pm,\circ}\|}_\infty$. In order to derive our results on radiation,
we have to assume certain a priori bounds on the corresponding solutions of (\ref{retRVMc}).
In particular, the latter have to exist globally in time.

\begin{assumption}\label{ex-assum-VM}
\begin{itemize}
\item[(a)] For each  $c\ge 1$ the system (\ref{retRVMc}) has a unique
solution $f^\pm\in C^2(\R\times\R^3\times\R^3)$, $E\in C^2(\R\times\R^3; \R^3)$,
$B\in C^2(\R\times\R^3; \R^3)$, satisfying the initial condition (\ref{indat}).
\item[(b)] There exists $P_1>0$ such that $f^\pm(t, x, p)=0$ for $|p|\ge P_1$
and all $c\ge 1$. In particular, $f^\pm(t, x, p)=0$ for $|x|\ge R_0+P_1|t|$ by (\ref{olan}).
\item[(c)] For every $T>0$, $R>0$, and $P>0$ there exists a constant $M_1(T, R, P)>0$ such that
\[ |\partial_t^{\alpha+1} f^\pm(t, x, p)|+|\partial_t^\alpha\nabla_x f^\pm(t, x, p)|\le M_1(T, R, P) \]
for $|t|\le T$, $|x|\le R$, $|p|\le P$, and $\alpha=0, 1$, uniformly in $c\geq 1$.
\end{itemize}
\end{assumption}
Note that none of the constants in Assumption~\ref{ex-assum-VM} may depend on $c$.
The constants
\[ 
R_0, P_0, S_0, P_1, M_1 
\]
from (\ref{daten2}) and Assumption~\ref{ex-assum-VM} are considered
to be the ``basic'' ones. Any other constant which appears in an estimate
is only allowed to depend on these. Checking the arguments from \cite{sc1,sc2},
it can be shown that Assumption~\ref{ex-assum-VM} holds at least for sufficiently ``small''
initial data $f^{\pm, \circ}$. A more precise investigation of the set of initial data
leading to solutions which satisfy Assumption~\ref{ex-assum-VM} is not part of this paper.
The main point we want to make here is that whenever Assumption~\ref{ex-assum-VM} is verified,
then the technique described below can be employed.

We will need estimates relating the solutions of (\ref{retRVMc})
to the corresponding Newtonian problem obtained in the limit 
$c \to \infty$. This sort of information usually goes under the name
of post-Newtonian approximation; see \cite{schaeff,bauku}. For this, one formally expands
the solutions in powers of $c^{-1}$ as
\begin{eqnarray*}
   f^\pm & = & f^\pm_0+c^{-1}f^\pm_1+c^{-2}f^\pm_2+\ldots, \\
   E & = & E_0+c^{-1}E_1+c^{-2}E_2+\ldots, \\
   B & = & B_0+c^{-1}B_1+c^{-2}B_2+\ldots,
\end{eqnarray*}
with coefficient functions $f^\pm_j$, $E_j$, and $B_j$ independent of $c$.
Moreover, by (\ref{hatp-def}),
\[
\hat{p}=p-(c^{-2}/2)p^2p+\ldots,\quad \gamma=1-(c^{-2}/2)p^2+\ldots .
\]
These expansions can be substituted into (\ref{retRVMc}),
and comparing coefficients at every order gives a sequence of equations for the coefficients.
The Newtonian limit of (\ref{retRVMc}) is given
by the plasma physics case of the Vlasov-Poisson system:
\begin{equation}\label{VPpl}
   \left.\begin{array}{c}\partial_t f^\pm_0+p\cdot\nabla_x f^\pm_0
   \pm E_0\cdot\nabla_p f^\pm_0=0, \\[2ex]
   \displaystyle E_0(t, x)= \int \frac{x-y}{|x-y|^3}\,\rho_0(t,y)\,dy, \\[2ex]
   \displaystyle\rho_0=\int (f^+_0-f^-_0)\,dp, \\[2ex]
   f^\pm_0(0, x, p)=f^{\pm, \circ}(x, p).\end{array}\quad\right\}\tag{VPpl}
\end{equation}
The following proposition addresses the well-known solvability properties of (\ref{VPpl}).
Clearly, $(f^+_0, f^-_0, E_0)$ is independent of $c$, and we refer to e.g.~\cite{schaeffer,lindner}
for the regularity of the solution.

\begin{proposition}\label{exf0f2-VM}
There are constants $R_2, P_2>0$, and for every $T>0$, $R>0$, and $P>0$,
there is a constant $M_2(T, R, P)>0$, with the following properties.
For initial data $f^{\pm,\circ}$ as above, there exists
a unique global solution $(f^\pm_0, E_0)$ of (\ref{VPpl}) so that
\begin{itemize}
\item[(a)] $f^\pm_0\in C^\infty(\R\times\R^3\times\R^3)$ and $E_0\in C^\infty(\R\times\R^3; \R^3)$,
\item[(b)] if $|t|\le 1$, then $f^\pm_0(t, x, p)=0$ for $|x|\ge R_2$ or $|p|\ge P_2$,
\item[(c)] if $|t|\le T$, $|x|\le R$, $|p|\le P$, and $\alpha=0, 1$, then
\[ |\partial_t^\alpha f^\pm_0(t, x, p)|+|\partial_t^\alpha E_0(t, x)|\le M_2(T, R, P). \]
\end{itemize}
\end{proposition}
For the approximation of solutions of (\ref{retRVMc}) by solutions
of (\ref{VPpl}), we state the following result without proof;
the result follows like the analogous one for (\ref{RVMc}), cf.~\cite{schaeff,bauku}.

\begin{proposition}\label{newton} Choose the constants $P_1>0$
and $M_1(T, R, P)>0$ according to Assumption~\ref{ex-assum-VM}.
Then for every $T>0$, $R>0$, and $P>0$ there are constants $M_3(T, R, P)>0$
and $M_4(T, R)>0$ with the following property. If $c\ge 2P_1$, let $(f^\pm, E, B)$ and $(f^\pm_0, E_0)$
denote the global solutions of (\ref{retRVMc}) and (\ref{VPpl})
provided by Assumption~\ref{ex-assum-VM} and Proposition~\ref{exf0f2-VM},
respectively, with initial data as above. Then
\begin{itemize}
\item[(a)] $|f^\pm(t, x, p)-f^\pm_0(t, x, p)|\le M_3(T, R, P)\,c^{-2}$
           for $|t|\le T$, $|x|\le R$, and $|p|\le P$,
\item[(b)] $|E(t, x)-E_0(t, x)|\le M_4(T, R)c^{-2}$ for $|t|\le T$ and $|x|\le R$,
\item[(c)] $|B(t, x)|\le M_4(T, R)\,c^{-1}$
for $|t|\le T$ and $|x|\le R$.
\end{itemize}
\end{proposition}
It is important to note that all the ``derived'' constants $R_2, P_2, M_2, M_3, M_4$
appearing above do only depend on the basic constants $R_0, P_0, S_0, P_1, M_1$.
We are now ready to state our first main result.

\begin{theorem}[Radiation for (\ref{retRVMc})]\label{main2}
Put $r_\ast=\max\{2(R_0+P_1), R_2\}$ and
\[ {\cal M}_{{\rm RVM}}=\{(t, r, c):\,r\ge 2r_\ast,\,c\ge 2P_1,
   \,|t-c^{-1}r|\le 1,\,r\ge c^3\}. \]
If $(t, r, c)\in {\cal M}_{{\rm RVM}}$, then with $r=|x|$, $\bar{x}=\frac{x}{|x|}$, and $u=t-c^{-1}|x|$,
\begin{equation}\label{absch-rad-VM}
   \Big|\bar{x}\cdot (B\times E)(t, x)+c^{-4}r^{-2}\,
   |\bar{x}\times\partial_t^2 {\cal D}(u)|^2\Big|
   \le A(c^{-5}r^{-2}+c^{-2}r^{-3}+c^{-1}r^{-4}),
\end{equation}
for a constant $A>0$ depending only on $R_0, P_0, S_0, P_1, M_1$.
In particular,
\begin{eqnarray}\label{xb09-VM}
   \frac{d}{dt}\,{\cal E}^{{\rm RVM}}_r(t)
   & = & \frac{c}{4\pi}\int_{|x|=r}\bar{x}\cdot (B\times E)(t, x)\,d\sigma(x)
   \nonumber \\
   & = & -\frac{2}{3c^3}\,|\partial_t^2 {\cal D}(u)|^2+{\cal O}(c^{-4}+c^{-1}r^{-1}+r^{-2})
\end{eqnarray}
for $(t, r, c)\in {\cal M}_{{\rm RVM}}$.
Here ${\cal E}^{{\rm RVM}}_r(t)=\int_{|x|\le r} e_{{\rm RVM}}(t, x)\,dx$,
see (\ref{ervm-def}), and
\[ {\cal D}(u)=\int x\,\rho_0(u, x)\,dx \]
denotes the dipole moment associated to the Vlasov-Poisson system (\ref{VPpl}).
\end{theorem}

\begin{remark}\label{remVM}{\rm (a) The condition $r\ge c^3$ in ${\cal M}_{{\rm RVM}}$
is not needed for the proof of (\ref{absch-rad-VM}) and (\ref{xb09-VM}).
It just guarantees that $c^{-2}r^{-3}\le c^{-5}r^{-2}$ and $c^{-1}r^{-1}\le c^{-4}$.
\smallskip

\noindent
(b) The same estimate (\ref{absch-rad-VM}) can be derived, possibly with a different constant $A$,
if the condition $|u|\le 1$ is replaced by $|u|\le u_0$ for some constant $u_0>0$.
\smallskip

\noindent
(c) As long as the constants $R_0, P_0, S_0, P_1, M_1$ remain independent of $c$,
one can also allow for $c$-dependent initial data $f^{\pm,\,\circ}_c$, both for (\ref{retRVMc})
and (\ref{VPpl}). However, in this case the functions $(f^\pm_0, E_0)$ become $c$-dependent, too.
For instance, in the particular case
\[ f^{\pm,\,\circ}_c=f^{\pm,\,\circ}_0+c^{-1}f^{\pm,\,\circ}_1+c^{-2}f^{\pm, \circ}_{r, c}\,, \]
with $f^{\pm,\,\circ}_0, f^{\pm,\,\circ}_1$, and $f^{\pm,\,\circ}_{r,c}$ satisfying suitable bounds
(independently of $c$ for $f^{\pm,\,\circ}_{r, c}$), Theorem~\ref{main2} remains valid,
if $f^\pm_0$ and $E_0$ are replaced by the approximations $\tilde{f}^\pm_0+c^{-1}\tilde{f}^\pm_1$
and $\tilde{E}_0+c^{-1}\tilde{E}_1$, respectively.
Here $(\tilde{f}^\pm_0, \tilde{E}_0)$ is the solution of (\ref{VPpl}) for the initial data
$f^{\pm,\,\circ}_0$, and $(\tilde{f}^\pm_1, \tilde{E}_1)$ solves the Vlasov-Poisson system
linearized about $(\tilde{f}^\pm_0, \tilde{E}_0)$, under the initial condition
$\tilde{f}^\pm_1(0)=f^{\pm,\,\circ}_1$.
\smallskip

\noindent
(d) In the case of one species only, say $f^{-,\,\circ}=0$, there is no dipole radiation,
since then $\partial_t^2{\cal D}=0$, cf.\ (\ref{dt2D=0}) below.
\smallskip

\noindent
(e) For spherically symmetric solutions there is again no dipole radiation.
In fact, if $\rho_0(t, -x)=\rho_0(t, x)$ for $x\in\R^3$, then ${\cal D}=0$ by symmetry.
}
\end{remark}
\medskip
The proof of Theorem~\ref{main2} is given in Section~\ref{Vmproof}.

\subsection{Monopole Radiation in the Vlasov-Nordstr\"om System}

If we set all physical constants (except the speed of light $c$) equal to unity,
then the Vlasov-Nordstr\"om system is given by
\begin{equation}\label{VNc}
   \left.\begin{array}{c}
   \partial_t f+\hat{p}\cdot\nabla_x f
   -\Big[(S\phi)p+c^2\gamma\nabla\phi\Big]\cdot\nabla_p f=4(S\phi)f, \\[1ex]
   (-\partial_t^2+c^2\Delta)\phi=4\pi\mu, \\[1ex]
   \displaystyle\mu=\int\gamma f\,dp,
   \end{array}\quad\right\}\tag{VNc}
\end{equation}
where we continue to use the notation from (\ref{hatp-def}),
and where $S=\partial_t+\hat{p}\cdot\nabla$. The matter distribution
is modeled through the nonnegative density function $f=f(t, x, p)$,
whereas the scalar function $\phi=\phi(t, x)$ describes the gravitational field.
We refer to \cite{calo1,calre1,andcalre,sc3} for the global existence of smooth 
solutions to (\ref{VNc}).
In analogy to the passage from (\ref{RVMc}) to (\ref{retRVMc}),
the solutions of (\ref{VNc}) that are isolated from incoming radiation
are the solutions of the retarded system
\begin{equation}\label{retVNc}
   \left.\begin{array}{c}
   \partial_t f+\hat{p}\cdot\nabla_x f
   -\Big[(S\phi)p+c^2\gamma\nabla\phi\Big]\cdot\nabla_p f=4(S\phi)f, \\[2ex]
   \displaystyle\phi(t, x)=-c^{-2}\int\mu(t-c^{-1}|y-x|, y)\,\frac{dy}{|y-x|}\,, \\[2ex]
   \displaystyle\mu=\int\gamma f\,dp,
   \end{array}\quad\right\}\tag{retVNc}
\end{equation}
which we call the retarded Vlasov-Nordstr\"om system.
We continue to make the standing hypotheses (\ref{daten1}) and (\ref{daten2})
for the initial data $f(0, x, p)=f^\circ(x, p)$ of (\ref{retVNc}).
A solution of (\ref{retVNc}) satisfies the relation
\begin{equation} \label{chardarst}
   f(t, x, p) = f^\circ (X(0, t, x, p), P(0, t, x, p)) e^{4 \phi(t, x)},
\end{equation}
where $s\mapsto (X(s,t,x,p),P(s,t,x,p))$ denotes the solution of
the characteristic system
\begin{equation}\label{charsyst}
   \dot x = \hat p,\quad\dot p = - (S\phi)p - c^2\gamma\nabla\phi,
\end{equation}
with $X(t, t, x, p)=x$ and $P(t, t, x, p)=p$. This implies that
as long as the solution exists
\[ 0 \leq f(t,x,p) \leq {\|f^\circ\|}_\infty\,; \]
note that $\phi\leq 0$. Concerning solutions of (\ref{retVNc}),
we make the following

\begin{assumption}\label{ex-assum}
\begin{itemize}
\item[(a)] For each  $c\ge 1$ the system (\ref{retVNc}) has a unique
solution $f\in C^2(\R\times\R^3\times\R^3)$, $\phi\in C^2(\R\times\R^3)$,
satisfying the initial condition $f(0, x, p)=f^\circ(x, p)$.
\item[(b)] There exists $P_1>0$ such that $f(t, x, p)=0$ for $|p|\ge P_1$
and all $c\ge 1$; by (\ref{chardarst}), (\ref{charsyst})
this implies that $f(t, x, p)=0$ for $|x|\ge R_0+P_1|t|$.
\item[(c)] For every $T>0$, $R>0$, and $P>0$ there exists
a constant $M_1(T, R, P)>0$ such that
\[ |\partial_t^\alpha f(t, x, p)|\le M_1(T, R, P) \]
for $|t|\le T$, $|x|\le R$, $|p|\le P$, and $\alpha=1, 2$.
In addition, for every $T>0$ and $R>0$ there exists a constant $M_1(T, R)>0$ such that
\[ |\phi(t, x)|+|\nabla\phi(t, x)|
   +|\partial_t\phi(t, x)|\le M_1(T, R) \]
for $|t|\le T$ and $|x|\le R$, uniformly in $c\geq 1$.
\end{itemize}
\end{assumption}
Again $R_0, P_0, S_0, P_1, M_1$ are considered to be the ``basic'' constants,
all other constants being derived from these. We remark that for ``small'' initial data
the existence of global-in-time solutions is shown in \cite{friedrich},
where also bounds on the solutions are obtained. It is reasonable to expect
that these solutions have the required regularity for smooth initial data,
cf.~\cite{lindner}, and that on compact time intervals
estimates as in Assumption~\ref{ex-assum}~(c) can be derived uniformly in $c$.
The crucial assumption is the bound on the momentum support in part (b),
which needs to be uniform in $c$ as well.

The Newtonian approximation for $c \to \infty$ of (\ref{retVNc})
is found by means of the formal expansion
\begin{eqnarray*}
   f & = & f_0+c^{-1}f_1+c^{-2}f_2+\ldots, \\
   \phi & = & \phi_0+c^{-1}\phi_1+c^{-2}\phi_2+c^{-3}\phi_3+c^{-4}\phi_4+\ldots, 
\end{eqnarray*}
see \cite{calee,bauer}. Thereby it is verified that this (lowest order)
Newtonian approximation of (\ref{retVNc}) is given
by the gravitational case of the Vlasov-Poisson system
\begin{equation}\label{VPgr}
   \left.\begin{array}{c}\partial_t f_0+p\cdot\nabla_x f_0
   -\nabla\phi_2\cdot\nabla_p f_0=0, \\[2ex]
   \displaystyle\phi_2(t, x)=-\int \frac{\rho_0(t, y)}{|x-y|}\,dy, \\[2ex]
   \displaystyle\rho_0=\int f_0\,dp, \\[2ex]
   f_0(0, x, p)=f^\circ(x, p).\end{array}\quad\right\}\tag{VPgr}
\end{equation}
The analogue of Proposition~\ref{exf0f2-VM} is valid for (\ref{VPgr}).
Note that $(f_0, \phi_2)$ is independent of $c$.

\begin{proposition}\label{exf0f2}
There are constants $R_2, P_2>0$, and for every $T>0$, $R>0$, and $P>0$,
there is a constant $M_2(T, R, P)>0$, with the following properties.
For initial data $f^\circ$ as above, there exists
a unique global solution $(f_0, \phi_2)$ of (\ref{VPgr}) so that
\begin{itemize}
\item[(a)] $f_0\in C^\infty(\R\times\R^3\times\R^3)$ and $\phi_2\in C^\infty(\R\times\R^3)$,
\item[(b)] if $|t|\le 1$, then $f_0(t, x, p)=0$ for $|x|\ge R_2$ or $|p|\ge P_2$,
\item[(c)] if $|t|\le T$, $|x|\le R$, $|p|\le P$, and $\alpha=0, 1, 2$, then
\[ |\partial_t^\alpha f_0(t, x, p)|+|\partial_t^{\alpha+1}\phi_2(t, x)|
   +|\partial_t^\alpha\nabla\phi_2(t, x)|\le M_2(T, R, P). \]
\end{itemize}
\end{proposition}
By \cite{bauer2}, we also have the following rigorous result concerning
the Newtonian limit of (\ref{retVNc}).

\begin{proposition}\label{darwin} Choose the constants $P_1>0$
and $M_1(T, R, P)>0$ according to Assumption~\ref{ex-assum}.
Then for every $T>0$, $R>0$, and $P>0$ there are constants $M_3(T, R, P)>0$
and $M_4(T, R)>0$ with the following properties.
If $c\ge 2P_1$, let $(f, \phi)$ and $(f_0, \phi_2)$ denote the global solutions
of (\ref{retVNc}) and (\ref{VPgr}) provided by Assumption~\ref{ex-assum}
and Proposition~\ref{exf0f2}, respectively, with initial data as above. Then
\begin{itemize}
\item[(a)] $|f(t, x, p)-f_0(t, x, p)|\le M_3(T, R, P)\,c^{-2}$
           for $|t|\le T$, $|x|\le R$, and $|p|\le P$,
\item[(b)] $|\nabla\phi(t, x)|\le M_4(T, R)c^{-2}$ for $|t|\le T$ and $|x|\le R$,
\item[(c)] $|\partial_t\phi(t, x)-c^{-2}\partial_t\phi_2(t, x)|
+|\nabla\phi(t, x)-c^{-2}\nabla\phi_2(t, x)|\le M_4(T, R)\,c^{-4}$
for $|t|\le T$ and $|x|\le R$.
\end{itemize}
\end{proposition}
After these preparations we can state our second main result.

\begin{theorem}[Radiation for (\ref{retVNc})]\label{main-gen}
Put $r_\ast=\max\{2(R_0+P_1), R_2\}$ and
\[ 
{\cal M}_{{\rm VN}}=\{(t, r, c):\,r\ge 2r_\ast,\,c\ge 2P_1,\,|t-c^{-1}r|\le 1,\,r\ge c^6\}. 
\]
If $(t, r, c)\in {\cal M}_{{\rm VN}}$, then with $r=|x|$, 
$\bar{x}=\frac{x}{|x|}$, and $u=t-c^{-1}|x|$,
\begin{equation}\label{absch-rad}
   \Big|\bar{x}\cdot (\partial_t\phi\nabla\phi)(t, x)
   +c^{-9}r^{-2}\big(\partial_t {\cal R}(\bar{x}, u)\big)^2\Big|
   \le A(c^{-10}r^{-2}+c^{-4}r^{-3}),
\end{equation}
for a constant $A>0$ depending only on $R_0, P_0, P_1, M_1, S_0$.
In particular,
\begin{eqnarray}\label{xb09}
   \frac{d}{dt}\,{\cal E}^{{\rm VN}}_r(t)
   & = & \frac{c^4}{4\pi}\int_{|x|=r}\bar{x}\cdot (\partial_t \phi\nabla\phi)(t, x)\,d\sigma(x)
   \nonumber \\ & = & -\frac{1}{4\pi c^5}\int_{|\omega|=1}
   \big(\partial_t {\cal R}(\omega, u)\big)^2\,d\sigma(\omega)
   +{\cal O}(c^{-6}+r^{-1})
\end{eqnarray}
for $(t, r, c)\in {\cal M}_{{\rm VN}}$.
Here ${\cal E}^{{\rm VN}}_r(t)=\int_{|x|\le r} e_{{\rm VN}}(t, x)\,dx$,
see (\ref{evn-def}), and
\begin{equation}\label{calR-def}
   {\cal R}(\bar{x}, u)=-\frac{1}{4\pi}\int|\bar{x}\cdot\nabla\phi_2(u, y)|^2\,dy
   -\int\!\!\!\int\,(\bar{x}\cdot p)^2 f_0(u, y, p)\,dp\,dy+4\,{\cal E}_{{\rm kin}}(u),
\end{equation}
where
\begin{equation}\label{ekin-def}
   {\cal E}_{{\rm kin}}(t)=\frac{1}{2}\int\!\!\!\int p^2 f_0(t, x, p)\,dp\,dx
\end{equation}
denotes the kinetic energy associated to the Vlasov-Poisson system (\ref{VPgr}).
\end{theorem}
Defining ${\cal E}_{{\rm pot}}(t)=-\frac{1}{8\pi}\int |\nabla\phi_2(t, x)|^2\,dx$,
the total energy ${\cal E}(t)={\cal E}_{{\rm kin}}(t)+{\cal E}_{{\rm pot}}(t)$
is conserved along solutions of (\ref{VPgr}).

\begin{remark}\label{rema}{\rm (a) Once again the condition $r\ge c^6$
in ${\cal M}_{{\rm VN}}$ is not needed for the proof of (\ref{absch-rad}).
It only has to be included in order that the second error term ${\cal O}(c^{-4}r^{-3})$
is at least as good as the first one, which is ${\cal O}(c^{-10}r^{-2})$.
\smallskip

\noindent
(b) In the sense of Remark~\ref{remVM}~(b) and (c),
one could allow for $|u|\le u_0$ and/or $c$-dependent initial data.
}
\end{remark}
\medskip
For spherically symmetric solutions, Theorem~\ref{main-gen} simplifies as follows.

\begin{cor}[Radiation for spherically symmetric solutions to (\ref{retVNc})]\label{main-spher}
\hspace{0em}\\ Define $r_\ast=\max\{2(R_0+P_1), R_2\}$ and
\[ 
{\cal M}_{{\rm VN}}=\Big\{(t, r, c):\,r\ge 2r_\ast,\,c\ge 2P_1,\,|t-c^{-1}r|\le 1,\,r\ge c^6\Big\}. 
\]
If $(t, r, c)\in {\cal M}_{{\rm VN}}$, then with $r=|x|$ and $u=t-c^{-1}r$,
\[ \Big|(\partial_t\phi\,\partial_r\phi)(t, x)
   +\frac{64}{9}\,c^{-9}r^{-2}\big(\partial_t {\cal E}_{{\rm kin}}(u)\big)^2\Big|
   \le A(c^{-10}r^{-2}+c^{-4}r^{-3}), \]
for a constant $A>0$ depending only on $R_0, P_0, S_0, P_1, M_1$. In particular,
\[ \frac{d}{dt}\,{\cal E}^{{\rm VN}}_r(t)
   =\frac{c^4}{4\pi}\int_{|x|=r}(\partial_t \phi\,\partial_r\phi)(t, x)\,d\sigma(x)
   =-\frac{64}{9c^5}\Big(\partial_t {\cal E}_{{\rm kin}}(u)\Big)^2
   +{\cal O}(c^{-6}+r^{-1}) \]
for $(t, r, c)\in {\cal M}_{{\rm VN}}$.
\end{cor}
The proofs of Theorem~\ref{main-gen} and Corollary~\ref{main-spher}
are carried out in Section~\ref{sect-gen}.

%%%%%%%%%%%%%%%%%%%%%%%%%%%%%%%%%%%%%%%%%%%%%%%%%%%%%%%%%%%%%%%%%%%%%%%%%%%%%%%%%%%%%%%%%%

\setcounter{equation}{0}

\section{Proofs}

\subsection{Proof of Theorem~\ref{main2}}
\label{Vmproof}

To expand $E(t, x)$ and $B(t, x)$ as given by (\ref{retRVMc}),
we recall from Assumption~\ref{ex-assum-VM}~(b) that $f^\pm(t, x, p)=0$
for $|x|\ge R_0+P_1|t|$. It follows that $\rho(t, x)=0$ and $j(t, x)=0$
for $|x|\ge R_0+P_1|t|$. If $(t, x, c)\in{\cal M}_{{\rm RVM}}$,
then $|u|=|t-c^{-1}|x||\le 1$ and $c\ge 2P_1$. Thus if $|y|\ge 2(R_0+P_1)$, then
\begin{eqnarray*}
   R_0+P_1 |t-c^{-1}|y-x|| & = & R_0+P_1 |u+c^{-1}|x|-c^{-1}|y-x||
   \\ & \le & R_0+P_1(|u|+c^{-1}|y|)\le R_0+P_1(1+(2P_1)^{-1}|y|)\le |y|.
\end{eqnarray*}
Hence $F(t-c^{-1}|y-x|, y)=0$ for both $F=-(\nabla\rho+c^{-2}\partial_t j)$
or $F=c^{-1}\nabla\times j$. Thus for the $y$-integrals defining $E$ and $B$
in (\ref{retRVMc}), it is sufficient to extend these over the ball 
$|y|\le\max\{2(R_0+P_1), R_2\}=r_\ast$.

In what follows $g={\cal O}(c^{-k}r^{-l})$ denotes a function such that
\[ |g(t, x)|\le Ac^{-k}r^{-l}\quad\text{for all}\quad |x|=r\ge 2 r_\ast,
   \quad c\ge 2P_1,\quad\text{and}\quad |t-c^{-1}|x||\le 1, \]
with $A$ only depending on the basic constants.
The following lemma states a representation for $E$ and $B$
similar to the Friedlander radiation field; see \cite[p.~91/92]{hoerm} and \cite{sc2}.

\begin{lemma}\label{friedl} The fields can be written as
\[ E(t, x)=E^{{\rm rad}}(t, x)+{\cal O}(r^{-2})\quad\mbox{and}\quad
   B(t, x)=B^{{\rm rad}}(t, x)+{\cal O}(c^{-1}r^{-2}), \]
where
\begin{eqnarray}
   E^{{\rm rad}}(t, x) & = & -r^{-1}\int_{|y|\le r_\ast}(\nabla\rho+c^{-2}\partial_t j)
   (u+c^{-1}\,\bar{x}\cdot y, y)\,dy, \label{Erad}
   \\[1ex] B^{{\rm rad}}(t, x) & = & c^{-1}r^{-1}\int_{|y|\le r_\ast}\nabla\times j
   (u+c^{-1}\,\bar{x}\cdot y, y)\,dy. \label{Brad}
\end{eqnarray}
\end{lemma}
{\bf Proof\,:} Consider $E$ first, and let $F=-(\nabla\rho+c^{-2}\partial_t j)$.
According to Assumption~\ref{ex-assum-VM}~(c), we have
$|F(\ldots)|\le AM_1(1+r_\ast, r_\ast, p_\ast)={\cal O}(1)$ for some constant $A>0$,
where $p_\ast=\max\{P_1, P_2\}$ and
\[ (\ldots)=(t-c^{-1}|y-x|, y)=(u+c^{-1}|x|-c^{-1}|y-x|, y). \]
If $|x|=r\ge 2r_\ast$ and $|y|\le r_\ast$, then $\frac{|x|}{|y-x|}
\le\frac{|x|}{|x|-r_\ast}\le 2$. It follows that
\[ \frac{1}{|y-x|}=\frac{1}{|x|}+\frac{|x|-|y-x|}{|y-x||x|}=r^{-1}+{\cal O}(r^{-2}) \]
for all $|y|\le r_\ast$. Therefore by (\ref{retRVMc}),
\begin{eqnarray*}
   E(t, x) & = & \int F(\ldots)\,\frac{dy}{|y-x|} =\int_{|y|\le r_\ast}\,F(\ldots)\frac{dy}{|y-x|}\,
   =\int_{|y|\le r_\ast} F(\ldots)\,\Big(r^{-1}+{\cal O}(r^{-2})\Big)\,dy
   \\ & = & r^{-1}\int_{|y|\le r_\ast} F(\ldots)\,dy+{\cal O}(r^{-2}).
\end{eqnarray*}
Next we note that for $|y|\le r_\ast$ and $|x|=r\ge 2r_\ast$,
\begin{eqnarray}\label{diff-expa}
   |x| - |x-y|
   & = & |x| - |x| \sqrt{1-2 \bar x \cdot y /|x| + |y|^2/|x|^2} \nonumber \\
   & = & |x| - |x| \left(1+\frac{1}{2} \left(-2 \bar x \cdot y/|x| + |y|^2/|x|^2\right)
   +{\cal O}(r^{-2}) \right) \nonumber \\
   & = & \bar x \cdot y + {\cal O}(r^{-1}).
\end{eqnarray}
Since
\[ |F(\ldots)-F(u+c^{-1}\,\bar{x}\cdot y, y)|
   \le {\|\partial_t F\|}_{L^\infty}c^{-1}||x|-|y-x|-\bar{x}\cdot y|
   ={\cal O}(c^{-1}r^{-1}) \]
by Assumption~\ref{ex-assum-VM}~(c) and (\ref{diff-expa}),
we get $E=E^{{\rm rad}}+{\cal O}(r^{-2})$. The proof for the magnetic field is analogous,
using $F=c^{-1}\nabla\times j$. {\hfill$\Box$}\bigskip

Now we need to investigate the relation between $E^{{\rm rad}}$ and $B^{{\rm rad}}$.
For this, we recall the continuity equation $\partial_t \rho+\nabla\cdot j = 0$
and calculate
\begin{eqnarray*}
   \nabla\rho(\ast) & = & \nabla_y\,[\rho(\ast)]+c^{-1}\bar{x}\,\nabla_y\cdot [j(\ast)]
   -c^{-2}(\bar{x}\cdot\partial_t j(\ast))\,\bar{x},
   \\ \nabla\times j(\ast) & = & \nabla_y\times [j(\ast)]-c^{-1}\bar{x}\times\partial_t j(\ast),
\end{eqnarray*}
where
\[ (\ast)=(u+c^{-1}\,\bar{x}\cdot y, y) \]
is the argument. This follows just from evaluating the total derivatives.
Since $\int_{|y|\le r_\ast} dy=\int dy$ in (\ref{Erad}) and (\ref{Brad})
by Assumption~\ref{ex-assum-VM}~(b), integration by parts shows that all $\nabla_y$-terms
drop out. Consequently, due to $u=t-c^{-1}r$ the relations
\begin{eqnarray}
   E^{{\rm rad}}(t, x) & = & -r^{-1}\int_{|y|\le r_\ast}[\nabla\rho(\ast)+c^{-2}\partial_t j(\ast)]\,dy
   \nonumber
   \\ & = & -r^{-1}\int_{|y|\le r_\ast}\Big[-c^{-2}(\bar{x}\cdot\partial_t j(\ast))\,\bar{x}
   +c^{-2}\partial_t j(\ast)\Big]\,dy
   \nonumber
   \\ & = & -c^{-2}r^{-1}\,\partial_t\int_{|y|\le r_\ast}\Big[j(u+c^{-1}\,\bar{x}\cdot y, y)
   -(\bar{x}\cdot j(u+c^{-1}\,\bar{x}\cdot y, y))\bar{x}\Big]\,dy, \label{kleli}
   \\ B^{{\rm rad}}(t, x) & = & c^{-1}r^{-1}\int_{|y|\le r_\ast}\nabla\times j(\ast)\,dy
   \nonumber
   \\ & = & -c^{-2}r^{-1}\,\partial_t\int_{|y|\le r_\ast}\bar{x}\times j(u+c^{-1}\,\bar{x}\cdot y, y)\,dy
   \nonumber
\end{eqnarray}
are obtained. Note that in particular $E^{{\rm rad}}$ and $B^{{\rm rad}}$ are of the same order
in $c^{-1}$ and $r^{-1}$, i.e.,
\begin{equation}\label{EBrad-ord}
   E^{{\rm rad}}(t, x)=B^{{\rm rad}}(t, x)={\cal O}(c^{-2}r^{-1})
\end{equation}
by Assumption~\ref{ex-assum-VM}~(c). Observing
\[ \bar{x}\times\int_{|y|\le r_\ast}\Big[j(\ast)-(\bar{x}\cdot j(\ast))\bar{x}\Big]\,dy
   =\int_{|y|\le r_\ast}\bar{x}\times j(\ast)\,dy, \]
differentiation w.r.~to $t$ yields the important formula
\begin{equation}\label{EBrad-bez}
   \bar{x}\times E^{{\rm rad}}(t, x)=B^{{\rm rad}}(t, x).
\end{equation}
Also
\[ \bar{x}\cdot\int_{|y|\le r_\ast}\Big[j(\ast)-(\bar{x}\cdot j(\ast))\bar{x}\Big]\,dy=0, \]
so that
\begin{equation}\label{ortho}
   \bar{x}\cdot E^{{\rm rad}}(t, x)=\bar{x}\cdot B^{{\rm rad}}(t, x)=0.
\end{equation}
Collecting the results from Lemma~\ref{friedl} and (\ref{EBrad-ord}), it follows that
\begin{eqnarray}\label{amkin}
   \bar{x}\cdot (B\times E) & = & \bar{x}\cdot\Big([B^{{\rm rad}}+{\cal O}(c^{-1}r^{-2})]
   \times [E^{{\rm rad}}+{\cal O}(r^{-2})]\Big)
   \nonumber \\ & = & \bar{x}\cdot\Big(B^{{\rm rad}}\times E^{{\rm rad}}
   +{\cal O}(c^{-2}r^{-3})+{\cal O}(c^{-3}r^{-3})+{\cal O}(c^{-1}r^{-4})\Big)
   \nonumber \\ & = & -|\bar{x}\times E^{{\rm rad}}|^2+{\cal O}(c^{-2}r^{-3})+{\cal O}(c^{-1}r^{-4}),
\end{eqnarray}
since by (\ref{EBrad-bez}) and (\ref{ortho}),
\begin{eqnarray*}
   \bar{x}\cdot (B^{{\rm rad}}\times E^{{\rm rad}}) & = & \bar{x}\cdot ([\bar{x}\times E^{{\rm rad}}]
   \times E^{{\rm rad}})=\bar{x}\cdot ((\bar{x}\cdot E^{{\rm rad}})E^{{\rm rad}}
   -|E^{{\rm rad}}|^2\bar{x}) \\ & = & -|E^{{\rm rad}}|^2=-|\bar{x}\times E^{{\rm rad}}|^2.
\end{eqnarray*}
Eqns.~(\ref{amkin}) and (\ref{kleli}) imply
\begin{equation}\label{holvor}
   \bar{x}\cdot (B\times E)=-c^{-4}r^{-2}\,\Big|\bar{x}\times\partial_t\int_{|y|\le r_\ast}
   j(u+c^{-1}\,\bar{x}\cdot y, y)\,dy\,\Big|^2
   +{\cal O}(c^{-2}r^{-3})+{\cal O}(c^{-1}r^{-4}).
\end{equation}
To expand the square as $c\to\infty$, we note that $|p|\ge p_\ast\ge P_1$ implies
$f^\pm(t, y, p)=0$ for all $t\in\R$ and all $y\in\R^3$ by Assumption~\ref{ex-assum-VM}~(b).
Therefore we can always replace the average over momentum space $\int\,dp$ by $\int_{|p|\le p_\ast}dp$.
For $|p|\le p_\ast$,
\[ \nabla_p\hat{p}=\gamma {\rm id}_{\R^3}-c^{-2}\gamma^3\,p\otimes p
   ={\rm id}_{\R^3}+{\cal O}(c^{-2}) \]
by (\ref{hatp-def}). Furthermore, using Assumption~\ref{ex-assum-VM},
\begin{eqnarray*}
   \nabla_x(f^+-f^-)(\ast) & = & \nabla_y[(f^+-f^-)(\ast)] - c^{-1}\bar{x}\,\partial_t(f^+-f^-)(\ast)
   \\ & = & \nabla_y[(f^+-f^-)(\ast)] + {\bf 1}_{\{|y|\le r_\ast,\,|p|\le p_\ast\}}{\cal O}(c^{-1}).
\end{eqnarray*}
Utilizing this, (\ref{retRVMc}), and Proposition~\ref{newton},
we get, writing $(\ast, p)=(u+c^{-1}\,\bar{x}\cdot y, y, p)$,
\begin{eqnarray}\label{dtj-rep}
   \int_{|y|\le r_\ast}\partial_t j(\ast)\,dy
   & = & \int\!\!\!\int\hat{p}\,\partial_t(f^+-f^-)(\ast, p)\,dp\,dy\nonumber \\
   & = & \int\!\!\!\int\hat{p}\,\bigg(-\hat{p}\cdot\nabla_x(f^+-f^-)(\ast, p)\nonumber \\
   & & \hspace{4.5em} -(E+c^{-1}\hat{p}\times B)\cdot \nabla_p(f^+ + f^-)(\ast, p)\bigg)\,dp\,dy
   \nonumber \\
   & = & {\cal O}(c^{-1})
   +\int_{|y|\le r_\ast}\int_{|p|\le p_\ast}\nabla_p\hat{p}\,(E+c^{-1}\hat{p}\times B)
   (f^+ + f^-)(\ast, p)\,dp\,dy \nonumber \\
   & = & {\cal O}(c^{-1}) \nonumber \\
   && {}+\int_{|y|\le r_\ast}\int_{|p|\le p_\ast}\big({\rm id}_{\R^3}+{\cal O}(c^{-2})\big)
   \big(E_0+{\cal O}(c^{-2})\big)\big(f^+_0+f^-_0+{\cal O}(c^{-2})\big)(\ast, p)\,dp\,dy
   \nonumber \\
   & = & \int_{|y|\le r_\ast}\int_{|p|\le p_\ast} E_0(f^+_0+f^-_0)(\ast, p)\,dp\,dy+{\cal O}(c^{-1}).
\end{eqnarray}
Also
\[ |E_0(f^+_0+f^-_0)(\ast, p)-E_0(f^+_0+f^-_0)(u, y, p)|
   \le {\|\partial_t\big(E_0(f^+_0+f^-_0)\big)\|}_{L^\infty}
   c^{-1}|\bar{x}\cdot y|={\cal O}(c^{-1})\]
by Proposition~\ref{exf0f2-VM}~(c). Thus (\ref{holvor}) and (\ref{dtj-rep}) yield
\begin{eqnarray}\label{kahl}
   \bar{x}\cdot (B\times E) & = & -c^{-4}r^{-2}\,
   \Big|\bar{x}\times\int_{|y|\le r_\ast}\int_{|p|\le p_\ast}
   E_0(f^+_0+f^-_0)(u, y, p)\,dp\,dy+{\cal O}(c^{-1})\Big|^2
   \nonumber \\ & & +\,{\cal O}(c^{-2}r^{-3})+{\cal O}(c^{-1}r^{-4})
   \nonumber \\ & = & -c^{-4}r^{-2}\,
   \Big|\bar{x}\times\int\!\!\!\int\,E_0(f^+_0+f^-_0)(u, y, p)\,dp\,dy\Big|^2
   \nonumber \\ & & +{\cal O}(c^{-5}r^{-2})+{\cal O}(c^{-2}r^{-3})+{\cal O}(c^{-1}r^{-4}),
\end{eqnarray}
since by Proposition~\ref{exf0f2-VM}~(b), $f^\pm_0(u, y, p)=0$ for $|y|\ge r_\ast\ge R_2$ or $|p|\ge p_\ast\ge P_2$. 
Defining the dipole moment
\[ {\cal D}(t)=\int x\,\rho_0(t, x)\,dx \]
with $\rho_0$ from (\ref{VPpl}), we obtain by the Vlasov equation in (\ref{VPpl}) that
\begin{eqnarray*}
   \partial_t {\cal D} & = & \int\!\!\!\int x\,\partial_t (f^+_0-f^-_0)\,dp\,dx
   =-\int\!\!\!\int x\,\big(p\cdot\nabla_x (f^+_0-f^-_0)
   +E_0\cdot\nabla_p (f^+_0+f^-_0)\big)\,dp\,dx \\
   & = &
   \int\!\!\!\int p\,(f^+_0-f^-_0)\,dp\,dx
\end{eqnarray*}
and
\begin{eqnarray}\label{heckr}
   \partial_t^2 {\cal D} & = & \int\!\!\!\int p\,\partial_t\,(f^+_0-f^-_0)\,dp\,dx
    =  -\int\!\!\!\int p\,\big(p\cdot \nabla_x(f^+_0-f^-_0)+E_0\cdot\nabla_p(f^+_0+f^-_0)\big)\,dp\,dx
   \nonumber \\
    & = & \int\!\!\!\int\,E_0(f^+_0+f^-_0)\,dp\,dx.
\end{eqnarray}
Due to (\ref{kahl}) it follows that
\[ \bar{x}\cdot (B\times E)=-c^{-4}r^{-2}\,
   |\bar{x}\times\partial_t^2 {\cal D}(u)|^2
   +{\cal O}(c^{-5}r^{-2})+{\cal O}(c^{-2}r^{-3})+{\cal O}(c^{-1}r^{-4}), \]
which completes the proof of (\ref{absch-rad-VM}).
Concerning (\ref{xb09-VM}), we have
\[ \frac{d}{dt}\,{\cal E}^{{\rm RVM}}_r(t)
   =\frac{c}{4\pi}\int_{|x|=r}\bar{x}\cdot (B\times E)(t, x)\,d\sigma(x)
   -c^2\int_{|x|=r} \int (\bar{x}\cdot p)(f^++f^-)(t, x, p)\, dp\,d\sigma(x) 
\]
by (\ref{ddtE-form}) and (\ref{Prvm-def}). For $|x|=r$ and $(t, r, c)\in {\cal M}_{{\rm RVM}}$,
we can estimate $R_0+P_1|t|=R_0+P_1|u+c^{-1}r|\le R_0+P_1+\frac{r}{2}\le r=|x|$,
since $r\ge 2r_\ast\ge 2(R_0+P_1)$. Therefore $f^\pm (t, x, p)=0$
by Assumption~\ref{ex-assum-VM}~(b), and this yields
\[ \frac{d}{dt}\,{\cal E}^{{\rm RVM}}_r(t)
   =\frac{c}{4\pi}\int_{|x|=r}\bar{x}\cdot (B\times E)(t, x)\,d\sigma(x),
   \quad (t, r, c)\in {\cal M}_{{\rm RVM}}. \]
Hence for (\ref{xb09-VM}) it suffices to use (\ref{absch-rad-VM}) and to note that
\[ -\frac{c^{-3}r^{-2}}{4\pi}\int_{|x|=r} |\bar{x}\times\partial_t^2 {\cal D}(u)|^2\,d\sigma(x)
   =-\frac{1}{4\pi c^3}\int_{|\omega|=1} |\omega\times\partial_t^2 {\cal D}(u)|^2\,d\sigma(\omega)
   =-\frac{2}{3c^3}\,|\partial_t^2 {\cal D}(u)|^2 \]
by integration. {\hfill$\Box$}\bigskip

\noindent
{\bf Proof of Remark~\ref{remVM}~(d):} If $f^-_0(t=0)=f^{-,\,\circ}=0$,
then also $f^-_0=0$ by (\ref{fpmdarst}). Thus defining
$\phi_0(t, x)=\int\!\!\!\int |x-y|^{-1}f^+_0(t, y, p)\,dp\,dy=\int |x-y|^{-1}\rho_0(t, y)\,dy$,
we get $E_0=-\nabla\phi_0$ and $\Delta\phi_0=-4\pi\rho_0$. Consequently, by (\ref{heckr}),
\begin{equation}\label{dt2D=0}
   \partial_t^2 {\cal D}=\int\!\!\!\int\,E_0 f^+_0\,dp\,dx=\int E_0\rho_0\,dx
   =\frac{1}{4\pi}\int\nabla\phi_0\,\Delta\phi_0\,dx=0.
\end{equation}
Hence there is no dipole radiation in this case.
{\hfill$\Box$}\bigskip

\subsection{Proof of Theorem~\ref{main-gen}}
\label{sect-gen}

By (\ref{retVNc}), $\partial_t \phi$ and $\nabla \phi$ are given by
\begin{eqnarray*}
  \partial_t \phi(t, x) & = & -c^{-2}\int \partial_t \mu(t-c^{-1}|y-x|, y)\,\frac{dy}{|y-x|}, \\
  \nabla\phi(t, x) & = & -c^{-2}\int \nabla\mu(t-c^{-1}|y-x|, y)\,\frac{dy}{|y-x|}.
\end{eqnarray*}
In full analogy to Lemma~\ref{friedl}, we obtain the following representation.

\begin{lemma} \label{grav-lem} We can write
\[ \partial_t \phi(t, x)=(\partial_t \phi)^{{\rm rad}}(t, x)+{\cal O}(c^{-2} r^{-2})
   \quad\mbox{and}\quad
   \nabla\phi(t, x)=(\nabla\phi)^{{\rm rad}}(t, x)+{\cal O}(c^{-2}r^{-2}), \]
where
\begin{eqnarray}
   (\partial_t\phi)^{{\rm rad}}(t, x) & = & -c^{-2} r^{-1}\int_{|y|\le r_\ast}
   \partial_t\mu(u+c^{-1}\,\bar{x}\cdot y, y)\,dy, \label{dtphi}
   \\[1ex] (\nabla\phi)^{{\rm rad}}(t, x) & = & -c^{-2}r^{-1}\int_{|y|\le r_\ast}
   \nabla\mu(u+c^{-1}\,\bar{x}\cdot y, y)\,dy. \label{nabphi}
\end{eqnarray}
\end{lemma}
Let again $(\ast)=(u+c^{-1}\,\bar{x}\cdot y, y)$ denote the argument.
Then $\nabla_y[\mu(\ast)]=c^{-1}\bar{x}\,\partial_t\mu(\ast)+\nabla\mu(\ast)$.
Since $\int_{|y|\le r_\ast} dy=\int\,dy$ in (\ref{nabphi}), it follows that
\[ \bar{x}\cdot(\nabla\phi)^{{\rm rad}}(t, x)
   =-c^{-2}r^{-1}\,\bar{x}\cdot\int
   \nabla\mu(\ast)\,dy=c^{-3}r^{-1}\,\bar{x}\cdot\int
   \bar{x}\,\partial_t\mu(\ast)\,dy=-c^{-1}(\partial_t\phi)^{{\rm rad}}(t, x). \]
The same argument shows that $(\nabla\phi)^{{\rm rad}} = {\cal O}(c^{-3}r^{-1})$,
and also $(\partial_t \phi)^{{\rm rad}}={\cal O}(c^{-2}r^{-1})$ by (\ref{dtphi}).
Hence we find from Lemma~\ref{grav-lem} that
\begin{equation}\label{rep-1}
   \bar{x}\cdot (\partial_t \phi\,\nabla\phi)(t, x)=-c^{-5}r^{-2}
   \,\bigg|\int\partial_t \mu(\ast)\,dy\bigg|^2+{\cal O}(c^{-4}r^{-3}).
\end{equation}
In order to expand the square we use, following \cite{glstr}, the differential operators
\[ 
T=c^{-1}\bar{x}\,\partial_t+\nabla\quad\text{and}\quad
   S=\partial_t +\hat{p}\cdot\nabla. 
\]
Then
\[\partial_t =(1-c^{-1}\hat{p}\cdot\bar{x})^{-1}(S-\hat{p}\cdot T) \]
and $\nabla_y [\mu(\ast)]=T\mu(\ast)$ is a total derivative. Hence the corresponding 
term drops out upon integration with respect to $y$. Observing the relation
\[ \nabla_p\cdot [(S\phi)p+c^2\gamma\nabla\phi]=3 (S\phi), \]
the Vlasov equation in (\ref{retVNc}) yields
\begin{eqnarray}\label{dtmu-rep}
   \int\partial_t\mu(\ast)\,dy & = & \int\!\!\!\int\gamma\,\partial_t f(\ast, p)\,dp\,dy
   \nonumber \\
   & = & \int\!\!\!\int\gamma\,(1-c^{-1}\hat{p}\cdot\bar{x})^{-1}
   \bigg([(S\phi)p+c^2\gamma\nabla\phi]\cdot\nabla_p f+4(S\phi)f\bigg)(\ast, p)\,dp\,dy
   \nonumber \\
   & = & -\int\!\!\!\int\nabla_p\big(\gamma\,(1-c^{-1}\hat{p}\cdot\bar{x})^{-1}\big)
   \cdot [(S\phi)p+c^2\gamma\nabla\phi]\,f(\ast, p)\,dp\,dy
   \nonumber \\
   & & +\int\!\!\!\int\gamma\,(1-c^{-1}\hat{p}\cdot\bar{x})^{-1}(S\phi)\,f(\ast, p)\,dp\,dy,
\end{eqnarray}
where $(\ast, p)=(u+c^{-1}\,\bar{x}\cdot y, y, p)$. A direct calculation shows that
\[
   \nabla_p\big(\gamma\,(1-c^{-1}\hat{p}\cdot\bar{x})^{-1}\big)
   = \nabla_p\big((\sqrt{1+p^2/c^2} - c^{-1} p\cdot\bar{x})^{-1}\big)
   = \gamma^2 (1-c^{-1}\hat{p}\cdot\bar{x})^{-2} (c^{-1} \bar{x} - c^{-2} \hat{p}).
\]
If $|p|\ge p_\ast\ge P_1$, then $f(\ast, p)=0$ by Assumption~\ref{ex-assum} (b).
Furthermore, if $|u|\le 1$ and $|x|\ge 2 r_\ast $, then $|y|\ge r_\ast$ enforces
$f(\ast, p)=0$ as before. Therefore we can replace $\int\!\!\!\int\,dp\,dy$
by $\int_{|y|\le r_\ast}\int_{|p|\le p_\ast} dy\,dp$ in the integrals occurring
in (\ref{dtmu-rep}). In other words, we may always assume that both $|y|$ and $|p|$
are bounded, with a bound depending only on the basic constants. Accordingly,
\begin{eqnarray}
   & & \gamma=1+{\cal O}(c^{-2}),\quad\gamma^2=1+{\cal O}(c^{-2}),
   \quad\hat{p}=\gamma p=p+{\cal O}(c^{-2}),
   \label{pker-ex1} \\
   & & (1-c^{-1}\hat{p}\cdot\bar{x})^{-1}=1+{\cal O}(c^{-1}),\quad
   (1-c^{-1}\hat{p}\cdot\bar{x})^{-2}=1+2 c^{-1} p\cdot\bar{x}+{\cal O}(c^{-2}).
   \label{pker-ex2}
\end{eqnarray}
This results in
\begin{equation}\label{nabp54}
   \nabla_p\big(\gamma\,(1-c^{-1}\hat{p}\cdot\bar{x})^{-1}\big)
   =c^{-1}\bar{x}+c^{-2}(2(p\cdot\bar{x})\bar{x}-p)+{\cal O}(c^{-3}).
\end{equation}
Furthermore, since $|u+c^{-1}\bar{x}\cdot y|\le 1+r_\ast$, also
\begin{eqnarray*}
   f(\ast, p) & = & f_0(\ast, p)+{\cal O}(c^{-2}), \\
   (S\phi)(\ast, p) & = & c^{-2}(\tilde{S}\phi_2)(\ast, p)+{\cal O}(c^{-4})
   = {\cal O}(c^{-2}), \\
   \nabla\phi(\ast) & = & c^{-2}\nabla\phi_2(\ast)+{\cal O}(c^{-4}),
\end{eqnarray*}
by Proposition~\ref{darwin} and (\ref{pker-ex1}),
where $\tilde{S}\phi_2=\partial_t\phi_2+p\cdot\nabla\phi_2$.
Observe that here the constants $M_3(1+r_\ast, r_\ast, p_\ast)$ and $M_4(1+r_\ast, r_\ast)$
enter the bounds on ${\cal O}(c^{-2})$ and ${\cal O}(c^{-4})$.
Hence from (\ref{dtmu-rep}), (\ref{pker-ex1}), (\ref{pker-ex2}), and (\ref{nabp54}) we get
\begin{eqnarray}\label{dtmu-rep2}
   \int\partial_t\mu(\ast)\,dy
   & = & -\int_{|y|\le r_\ast}\int_{|p|\le p_\ast}
   \Big(c^{-1}\bar{x}+c^{-2}(2(p\cdot\bar{x})\bar{x}-p)+{\cal O}(c^{-3})\Big)
   \nonumber \\ & & \hspace{5em}\cdot\Big[{\cal O}(c^{-2})
   +(1+{\cal O}(c^{-2}))\Big(\nabla\phi_2+{\cal O}(c^{-2})\Big)\Big]
   \Big(f_0(\ast, p)+{\cal O}(c^{-2})\Big)\,dp\,dy
   \nonumber \\
   & & +\int_{|y|\le r_\ast}\int_{|p|\le p_\ast}
   \Big(1+{\cal O}(c^{-2})\Big)\,\Big(1+{\cal O}(c^{-1})\Big)\Big(c^{-2}(\tilde{S}\phi_2)
   +{\cal O}(c^{-4})\Big)
   \nonumber \\ & & \hspace{6.5em}\Big(f_0(\ast, p)+{\cal O}(c^{-2})\Big)\,dp\,dy
   \nonumber \\ & = & -c^{-1}\bar{x}\cdot \int_{|y|\le r_\ast}\int_{|p|\le p_\ast}(1+2c^{-1}p\cdot\bar{x})
   \nabla (\phi_2 f_0)(\ast, p)\,dp\,dy \nonumber \\
   & & + c^{-2}\int_{|y|\le r_\ast}\int_{|p|\le p_\ast}
   (\tilde{S}\phi_2+p\cdot\nabla\phi_2)f_0(\ast, p)\,dp\,dy+{\cal O}(c^{-3}).
\end{eqnarray}
Let $\psi$ denote either $\nabla\phi_2$ or $\partial_t\phi_2$.
Then by Proposition~\ref{exf0f2}~(c),
\begin{eqnarray*}
   (\psi f_0)(\ast, p) & = & (\psi f_0)(u+c^{-1}\,\bar{x}\cdot y, y, p)
   \\ & = & (\psi f_0)(u, y, p)+c^{-1}(\bar{x}\cdot y)\,\partial_t (\psi f_0)(u, y, p)+{\cal O}(c^{-2})
   \\ & = & (\psi f_0)(u, y, p)+{\cal O}(c^{-1}).
\end{eqnarray*}
Hence (\ref{dtmu-rep2}) yields
\begin{eqnarray}\label{dtmu-rep3}
   \int\partial_t\mu(\ast)\,dy
   & = & -c^{-1}\bar{x}\cdot \int_{|y|\le r_\ast}\int_{|p|\le p_\ast}
   \Big((\nabla\phi_2 f_0)(u, y, p)+c^{-1}(\bar{x}\cdot y)\,\partial_t (\nabla\phi_2 f_0)(u, y, p)
   +{\cal O}(c^{-2})
   \nonumber \\ & & \hspace{10em} +2c^{-1} (\bar{x}\cdot p)\,\nabla\phi_2 f_0(u, y, p)\Big)\,dp\,dy
   \nonumber \\ & & +c^{-2}\int_{|y|\le r_\ast}\int_{|p|\le p_\ast}
   (\tilde{S}\phi_2+p\cdot\nabla\phi_2)f_0(u, y, p)\,dp\,dy+{\cal O}(c^{-3})
   \nonumber \\  & = &
   {\cal O}(c^{-3})-c^{-1}\int_{|y|\le r_\ast}(\bar{x}\cdot\nabla\phi_2)\,\rho_0(u, y)\,dy
   \nonumber \\
   & & -c^{-2}\partial_t\,\int_{|y|\le r_\ast}(\bar{x}\cdot y)\,(\bar{x}\cdot\nabla\phi_2)\,\rho_0(u, y)\,dy
   \nonumber \\
   & & +c^{-2}\int_{|y|\le r_\ast}\int_{|p|\le p_\ast}\Big(\tilde{S}\phi_2
   +p\cdot\nabla\phi_2-2(\bar{x}\cdot p)\,(\bar{x}\cdot\nabla\phi_2)\Big)
   f_0(u, y, p)\,dp\,dy,
\end{eqnarray}
recalling that $u=t-c^{-1}r$. In view of Proposition~\ref{exf0f2}~(b)
we may extend all integrals over the whole space again.
Now
\[ \int\nabla\phi_2\rho_0(u, y)\,dy
   =\int\!\!\!\int \nabla\phi_2(u, y)f_0(u, y, p)\, dp\,dy =0 \]
by Lemma~\ref{vp-props} below, whence the lowest order term drops out.
In addition, Lemma~\ref{vp-props} also shows that
\[ \int\!\!\!\int (\tilde{S}\phi_2)(u, y)f_0(u, y, p)\, dp\,dy
   =-2\,\partial_t {\cal E}_{{\rm kin}}(u) \]
as well as
\[ \int\!\!\!\int (p\cdot\nabla\phi_2)(u, y)f_0(u, y, p)\, dp\,dy
   =-\partial_t {\cal E}_{{\rm kin}}(u) \]
and
\[ \int\!\!\!\int (\bar{x}\cdot p)(\bar{x}\cdot\nabla\phi_2)(u, y)f_0(u, y, p)\, dp\,dy
   =-\frac{1}{2}\,\partial_t\int\!\!\!\int (\bar{x}\cdot p)^2 f_0(u, y, p)\,dp\,dy. \]
Finally, we can also write
\[ \int (\bar{x}\cdot y)(\bar{x}\cdot\nabla\phi_2)(u, y)\rho_0(u, y)\,dy
   =-{\cal E}_{{\rm pot}}(u)-\frac{1}{4\pi}\int|\bar{x}\cdot\nabla\phi_2(u, y)|^2\,dy \]
by Lemma~\ref{vp-props} and since $|\bar{x}|=1$. Using this and $\partial_t {\cal E}_{{\rm pot}}
=-\partial_t{\cal E}_{{\rm kin}}$ (see the remarks following (\ref{ekin-def}))
in (\ref{dtmu-rep3}), and collecting all the terms, it follows that
\begin{equation}\label{dtmu-rep4}
  \int\partial_t\mu(\ast)\,dy = -c^{-2}\,\partial_t {\cal R}(\bar{x}, u)+{\cal O}(c^{-3}),
\end{equation}
with ${\cal R}(\bar{x},u)$ as in (\ref{calR-def}).
Inserting (\ref{dtmu-rep4}) into (\ref{rep-1}), we see that
\begin{eqnarray*}
   \bar{x}\cdot (\partial_t\phi\nabla\phi)(t, x)
   & = & -c^{-5}r^{-2}\big|-c^{-2}\partial_t {\cal R}(\bar{x}, u)
   +{\cal O}(c^{-3})\big|^2+{\cal O}(c^{-4}r^{-3}) \\
   & = & -c^{-9}r^{-2}\big(\partial_t {\cal R}(\bar{x},u)\big)^2+{\cal O}(c^{-10}r^{-2})
   +{\cal O}(c^{-4}r^{-3}).
\end{eqnarray*}
Therefore (\ref{absch-rad}) is proved. Concerning (\ref{xb09}), the fact that
\[ \frac{d}{dt}\,{\cal E}^{{\rm VN}}_r(t)
   =\frac{c^4}{4\pi}\int_{|x|=r}\bar{x}\cdot (\partial_t \phi\nabla\phi)(t, x)\,d\sigma(x) \]
is due to $(t, r, c)\in {\cal M}_{{\rm VN}}$, analogously to the argument in the proof
of Theorem~\ref{main2}. Hence (\ref{xb09}) is a direct consequence of (\ref{absch-rad}),
changing variables as $x=r\omega$. {\hfill$\Box$}\bigskip

We still need to give the proof of

\begin{lemma}\label{vp-props} For the Vlasov-Poisson system (\ref{VPgr}),
\begin{eqnarray*}
   & & \int\!\!\!\int\nabla\phi_2(t, x)f_0(t, x, p)\,dp\,dx=0,\\
   & & \int\!\!\!\int (\tilde{S}\phi_2)(t, x)
   f_0(t, x, p)\,dp\,dx=-2\,\partial_t {\cal E}_{{\rm kin}}(t),
   \\ & & \int\!\!\!\int p\cdot\nabla\phi_2(t, x)f_0(t, x, p)\,dp\,dx=-\partial_t {\cal E}_{{\rm kin}}(t),
   \\ & & \int\!\!\!\int (\xi\cdot p)(\xi\cdot\nabla\phi_2)(t, x)f_0(t, x, p)\,dp\,dx
   =-\frac{1}{2}\,\partial_t\int\!\!\!\int (\xi\cdot p)^2 f_0(t, x, p)\,dp\,dx\quad (\xi\in\R^3),
   \\ & & \int\!\!\!\int (\xi\cdot x)(\xi\cdot\nabla\phi_2)(t, x)f_0(t, x, p)\,dp\,dx
   =-|\xi|^2\,{\cal E}_{{\rm pot}}(t)-\frac{1}{4\pi}\int|\xi\cdot\nabla\phi_2(t, x)|^2\,dx\quad(\xi\in\R^3),
\end{eqnarray*}
where  ${\cal E}_{{\rm kin}}(t)=\frac{1}{2}\int\!\!\!\int p^2 f_0(t, x, p)\,dp\,dx$,
see (\ref{ekin-def}), and ${\cal E}_{{\rm pot}}(t)=-\frac{1}{8\pi}\int |\nabla\phi_2(t, x)|^2\,dx$.
\end{lemma}
{\bf Proof\,:} Firstly, since $\Delta \phi_2 = 4 \pi \rho_0$,
\begin{eqnarray*}
   \int\!\!\!\int\nabla\phi_2(t, x)f_0(t, x, p)\,dp\,dx
   & = & \int \nabla\phi_2(t, x)\,\rho_0(t, x)\,dx 
     = \frac{1}{4 \pi}\int\nabla\phi_2(t, x)\,\Delta \phi_2 (t,x)\,dx\\
   & = & \frac{1}{8 \pi}\int\nabla\cdot|\nabla\phi_2(t, x)|^2\,dx = 0.
\end{eqnarray*}
For the remaining assertions we define the mass current density as $j_0=\int p f_0\,dp$.
Integration of the Vlasov equation with respect to $p$ implies the continuity equation
$\partial_t\rho_0+\nabla\cdot j_0=0$. Hence
\begin{eqnarray*}
   \int\!\!\!\int \tilde S \phi_2\,f_0 \,dp\,dx
   &=& \int (\partial_t \phi_2\,\rho_0 + \nabla\phi_2 \cdot j_0)\,dx
   =\int (\partial_t \phi_2\,\rho_0 - \phi_2\,\nabla \cdot j_0) \,dx
   \\ &=& \partial_t \int \phi_2 \rho_0\, dx = 2\partial_t {\cal E}_{{\rm pot}}(t)
   = -2\partial_t {\cal E}_{{\rm kin}}(t)
\end{eqnarray*}
by conservation of energy, and
\begin{eqnarray*}
   \int\!\!\!\int p\cdot\nabla\phi_2\,f_0\,dp\,dx
   & = & \int j_0\cdot \nabla \phi_2\,dx=\int \partial_t \rho_0\,\phi_2\, dx\\
   & = & -\int\!\!\!\int\frac{1}{|x-y|}\,\partial_t \rho_0(t,x) \rho_0 (t,y)\,dx\,dy \\
   & = & \partial_t {\cal E}_{{\rm pot}}(t) = - \partial_t {\cal E}_{{\rm kin}}(t).
\end{eqnarray*}
Furthermore, by (\ref{VPgr}),
\begin{eqnarray*}
   \partial_t\int\!\!\!\int (\xi\cdot p)^2 f_0\,dp\,dx
   & = & \int\!\!\!\int (\xi\cdot p)^2\,\nabla_p\cdot(\nabla\phi_2\,f_0)\,dp\,dx
   \\ & = & -2\int\!\!\!\int (\xi\cdot p)(\xi\cdot\nabla\phi_2)\,f_0\,dp\,dx.
\end{eqnarray*}
For the last assertion, using $\Delta\phi_2=4\pi\rho_0$,
\begin{eqnarray*}
   \int (\xi\cdot x)(\xi\cdot\nabla\phi_2)\rho_0\,dx
   & = & \frac{1}{4\pi}\sum_{i,j=1}^3\int(\xi\cdot x)
   \,\xi_i\partial_i\phi_2\,\partial_j\partial_j\phi_2\,dx
   \\ & = & -\frac{1}{4\pi}\sum_{i,j=1}^3
   \int\Big((\xi\cdot x)\xi_i\,\partial_i\partial_j\phi_2
   +\xi_j\xi_i\,\partial_i\phi_2\Big)\partial_j\phi_2\,dx
   \\ & = & -\frac{1}{8\pi}\int(\xi\cdot x)
   \xi\cdot\nabla|\nabla\phi_2|^2\,dx
   -\frac{1}{4\pi}\int|\xi\cdot\nabla\phi_2|^2\,dx
   \\ & = & \frac{|\xi|^2}{8\pi}\int|\nabla\phi_2|^2\,dx
   -\frac{1}{4\pi}\int|\xi\cdot\nabla\phi_2|^2\,dx
   \\ & = & -|\xi|^2\,{\cal E}_{{\rm pot}}-\frac{1}{4\pi}\int|\xi\cdot\nabla\phi_2|^2\,dx.
\end{eqnarray*}
This completes the proof of the lemma. {\hfill$\Box$}\bigskip

\subsection{Proof of Corollary~\ref{main-spher}}

In this section we verify Corollary~\ref{main-spher}
by specializing Theorem~\ref{main-gen} to spherically symmetric functions.
We recall that initial data $f^\circ$ are said to be
spherically symmetric, if
\[ f^\circ(Ax, Ap)=f^\circ(x, p) \]
for any matrix $A\in {\rm SO}(3)$. Then the solution $(f_0, \phi_2)$ of (\ref{VPgr})
provided by Proposition~\ref{exf0f2} remains spherically symmetric for all times.
Therefore
\begin{equation}\label{symm}
   f_0(t, Ax, Ap)=f_0(t, x, p),\quad\rho_0(t, Ax)=\rho_0(t, x),
   \quad\mbox{and}\quad\phi_2(t, x)=\phi_2(t, Ax)
\end{equation}
holds for all $A\in {\rm SO}(3)$.

Firstly, this implies $\nabla\phi_2=\bar{x}\,\partial_r\phi_2$
as well as $|\nabla\phi_2|^2=|\partial_r\phi_2|^2$, $\partial_r$ denoting the radial derivative.
By choosing $A\in {\rm SO}(3)$ such that $A\bar{x}=e_j$ (the $j$'s unit vector in $\R^3$),
(\ref{symm}) yields
\begin{eqnarray*}
   \frac{1}{4\pi}\int|\bar{x}\cdot\nabla\phi_2(u, y)|^2\,dy
   & = & \frac{1}{4\pi}\int|\partial_j\phi_2(u, y)|^2\,dy
   =\frac{1}{4\pi}\int\Big|\frac{y_j}{|y|}\,\partial_r\phi_2(u, y)\Big|^2\,dy
   \\ & = & \frac{1}{12\pi}\int|\partial_r\phi_2(u, y)|^2\,dy
   =-\frac{2}{3}\,{\cal E}_{{\rm pot}}(u).
\end{eqnarray*}
Similarly, (\ref{symm}) and $|\bar{x}|^2=1$ implies that
\begin{eqnarray*}
   \int\!\!\!\int\,(\bar{x}\cdot p)^2 f_0(u, y, p)\,dp\,dy
   & = & \int\!\!\!\int\,p_j^2\,f_0(u, y, p)\,dp\,dy=\frac{1}{3}\int\!\!\!\int\,p^2\,f_0(u, y, p)\,dp\,dy
   \\ & = & \frac{2}{3}\,{\cal E}_{{\rm kin}}(u).
\end{eqnarray*}
Therefore by (\ref{calR-def}),
\begin{eqnarray*}
   {\cal R}(\bar{x}, u) & = & -\frac{1}{4\pi}\int|\bar{x}\cdot\nabla\phi_2(u, y)|^2\,dy
   -\int\!\!\!\int\,(\bar{x}\cdot p)^2 f_0(u, y, p)\,dp\,dy + 4\,{\cal E}_{{\rm kin}}(u)
   \\ & = & \frac{2}{3}\,{\cal E}_{{\rm pot}}(u)+\frac{10}{3}\,{\cal E}_{{\rm kin}}(u).
\end{eqnarray*}
Since $\partial_t {\cal E}_{{\rm pot}}=-\partial_t {\cal E}_{{\rm kin}}$
by conservation of energy, we get $\partial_t {\cal R}(\bar{x}, u)
=\frac{8}{3}\,\partial_t {\cal E}_{{\rm kin}}(u)$.
Hence Corollary~\ref{main-spher} follows from (\ref{absch-rad}) and (\ref{xb09}).
{\hfill$\Box$}

\end{document}